%% file: main.tex
\documentclass[manuscript, screen, nonacm]{acmart}

\renewcommand\footnotetextcopyrightpermission[1]{} 
\usepackage{pgfplots}
\usepackage{siunitx}
\usepackage{algorithm}
\usepackage{algorithmic}
\usepackage{pgfplotstable}
\usepgfplotslibrary{statistics}
\usepackage{multirow}
\usepackage{collcell}
\usepackage{colortbl}
\usepackage{xcolor}
\usepackage{datatool}
\usepackage[linesnumbered,ruled,vlined,algo2e]{algorithm2e}
\usepackage{enumitem}

\SetCommentSty{mycommfont}

\usepgfplotslibrary{statistics}
\pgfplotsset{compat=newest}
\usepgfplotslibrary{fillbetween}
\usepgfplotslibrary{groupplots}

\pgfplotsset{
    compat=newest,
    /pgfplots/ybar legend/.style={
        /pgfplots/legend image code/.code={
            \draw[##1,/tikz/.cd,bar width=3pt,yshift=-0.2em,bar shift=0pt] plot coordinates {(0cm,0.8em)};
        },
    },
}

\newcommand\nissan{Nissan}
\newcommand\blue[1]{\textcolor{black}{#1}}

\begin{document}


\title{Online Decision-Making Under Uncertainty for Vehicle-to-Building Systems}

\author{Rishav Sen} 
\affiliation{%
  \institution{Vanderbilt University} 
  \city{Nashville} 
  \country{USA}
} 
\email{rishav.sen@vanderbilt.edu}

\author{Yunuo Zhang} 
\affiliation{%
  \institution{Vanderbilt University} 
  \city{Nashville} 
  \country{USA}
} 
\email{yunuo.zhang@vanderbilt.edu}

\author{Fangqi Liu} 
\affiliation{%
  \institution{Vanderbilt University} 
  \city{Nashville} 
  \country{USA}
} 
\email{fangqi.liu@vanderbilt.edu}

\author{Jose Paolo Talusan} 
\affiliation{%
  \institution{Vanderbilt University} 
  \city{Nashville} 
  \country{USA}
} 
\email{jose.paolo.talusan@vanderbilt.edu}

\author{Ava Pettet} 
\affiliation{%
  \institution{Nissan Advanced Technology Center - Silicon Valley} 
  \city{Santa Clara} 
  \country{USA}
} 
\email{ava.pettet@nissan-usa.com}

\author{Yoshinori Suzue} 
\affiliation{%
  \institution{Nissan Advanced Technology Center - Silicon Valley} 
  \city{Santa Clara} 
  \country{USA}
} 
\email{yoshinori.suzue@nissan-usa.com}

\author{Ayan Mukhopadhyay} 
\affiliation{%
  \institution{Vanderbilt University} 
  \city{Nashville} 
  \country{USA}
} 
\email{ayan.mukhopadhyay@vanderbilt.edu}

\author{Abhishek Dubey} 
\affiliation{%
  \institution{Vanderbilt University} 
  \city{Nashville} 
  \country{USA}
} 
\email{abhishek.dubey@vanderbilt.edu}

\renewcommand{\shortauthors}{Sen et al.}

\begin{abstract} 
\input{sections/00_abstract} 
\end{abstract}

\keywords{Monte Carlo Tree Search (MCTS), Optimization, Vehicle-to-Building (V2B), Electric Vehicle Charging}

\maketitle

\section{Introduction} 
\input{sections/01_introduction}

\section{Related Work} 
\input{sections/03_related_work}

\section{Problem Description} 
\input{sections/02_problem}

\input{sections/02b_mdp}

\section{Online Approach} 
\input{sections/04b_online_approach} 
\input{sections/04c_MCTS}

\section{Experiments and Analysis} 
\input{sections/05_experiment}

\section{Conclusion} 
\input{sections/06_conclusion}

\section{Acknowledgement} 
\input{sections/09_ack}

\bibliographystyle{ACM-Reference-Format} 
\bibliography{main}

\clearpage 
\appendix 
\input{sections/07_appendix}

\end{document}

%% file: sections/00_abstract.tex
Vehicle-to-building (V2B) systems integrate physical infrastructures, such as smart buildings and electric vehicles (EVs) connected to chargers at the building, with digital control mechanisms to manage energy use. By utilizing EVs as flexible energy reservoirs, buildings can dynamically charge and discharge them to optimize energy use and cut costs under time-variable pricing and demand charge policies. This setup leads to the V2B optimization problem, where buildings coordinate EV charging and discharging to minimize total electricity costs while meeting users' charging requirements.
However, the V2B optimization problem is challenging because of: (1) fluctuating electricity pricing, which includes both energy charges (\$/kWh) and demand charges (\$/kW); (2) long planning horizons (typically over 30 days); (3) heterogeneous chargers with varying charging rates, controllability, and directionality (i.e., unidirectional or bidirectional); and (4) user-specific battery levels at departure to ensure user requirements are met. In contrast to existing approaches that often model this setting as a single-shot combinatorial optimization problem, we highlight critical limitations in prior work and instead model the V2B optimization problem as a Markov decision process (MDP), i.e., a stochastic control process. Solving the resulting MDP is challenging due to the large state and action spaces. To address the challenges of the large state space, we leverage online search, and we counter the action space by using domain-specific heuristics to prune unpromising actions. 
We validate our approach in collaboration with
Nissan Advanced Technology Center - Silicon Valley. Using data from their EV testbed, we
show that the proposed framework significantly outperforms state-of-the-art methods.

%% file: sections/01_introduction.tex

Electric vehicles (EVs) are leading the shift in the global energy landscape towards more sustainable energy solutions~\citep{nykvist2015rapid}. Managing EV energy requirements presents both opportunities and challenges. In this work, we focus on one such opportunity---vehicle-to-building charging (V2B)---that involves co-optimizing the energy management of EVs and smart buildings. The key idea behind V2B charging exploits the ability to control the rates of charging and leverages the use of bidirectional EVs as energy storage facilities to add resilience and demand-response capabilities to smart buildings~\citep{doe}. For example, EVs can be charged when energy is available at lower costs and discharged to supply energy to buildings when energy costs are high, thereby promoting optimal energy use, and reducing peak power demands (highest instantaneous power usage in the billing period), substantially decreasing energy and demand costs for the building~\citep{sun2013peak}. A portion of the savings can then be shared with EV owners, either directly or through discounted charging, thereby producing a \textit{win-win} framework for both the building and the EV owners.

Despite the apparent simplicity of the V2B framework, operationalizing it presents several challenges. While EV owners can be incentivized to participate in such programs by offering charges at low (or zero) cost, strategies that solely optimize energy costs can result in arbitrarily low EV charges when the owner leaves the smart building, thereby causing inconvenience to the building owners. Therefore, a V2B optimization framework must ensure that EV owners depart with a pre-specified level of charge at their departure time while dealing with the exogenous uncertainties of fluctuating building load, EV arrivals (and departures), and energy prices. Since many EVs arrive throughout the day and charging requirements differ among vehicles, the optimization problem becomes highly complex. Modern buildings also use heterogeneous EV chargers (of different makes, models, rates of charging, and directionality), which further complicates the sequential decision-making process under uncertainty. Additionally, the V2B framework must be able to accommodate complex pricing policies of power companies, e.g., time-of-use energy charges (dollar per kilowatt-hours (kWh)) and demand charges (dollar per kilowatts (kW)); crucially, the demand cost is computed over a relatively longer temporal horizon, e.g., for most American power companies, this involves computing the maximum instantaneous power utilization over a month.

The underlying optimization problem for the V2B problem has been explored in prior work; e.g., \citet{tanguy2016optimization} present one of the most comprehensive optimization frameworks for this problem setting. Their model is elegant and extremely tractable---the decision problem (i.e., deciding \textit{when} and by \textit{how much} each vehicle is charged or discharged) is modeled as a linear program that can be efficiently solved in polynomial time, providing scalability by design. However, the inherent scalability comes at the cost of several assumptions that are not valid in practice: 1) the \textit{single-shot} optimization approach assumes that car arrivals and departures are known apriori, and computing all vehicle-to-charger assignments at once prohibits adaptability to dynamic changes, e.g., varying energy prices or building loads; 2) the linearity of the formulation implicitly relies on homogeneous charger configurations; and 3) the formulation captures only energy cost but cannot accommodate demand cost, which is an important determinant of V2B policies in practice. We address these limitations by formulating the V2B problem as a stochastic control process and modeling it as a Markov decision process (MDP). Our formulation is motivated by the observation that, in practice, decisions about charging and discharging vehicles (and by how much) must be taken sequentially under exogenous sources of uncertainty.

\label{mcts_diff}
However, computing an optimal policy for the MDP is very challenging in our problem setting, particularly due to \textit{complex credit assignment} and the \textit{curse of dimensionality}~\citep{sutton2018reinforcement}. Specifically, given that the decisions must be computed at a relatively high frequency (e.g., every 15 minutes) and the demand cost is only observed at the end of a billing period (usually a month), the decision maker's actions have long-term consequences, but the rewards or feedback may only be received much later. Moreover, as the planning horizon increases, the number of possible states and actions grow exponentially. This growth leads to a combinatorial explosion, making it computationally infeasible to consider all possible sequences of actions and states. To tackle these challenges, we propose an online approach that focuses on computing near-optimal actions for the current state of the system instead of seeking to learn a policy. Our approach is based on Monte Carlo tree search (MCTS)~\citep{kocsis2006bandit}, a general-purpose online algorithm for stochastic control processes. 

While MCTS tackles some of the challenges of the V2B setting by using a bandit-based strategy for exploring promising trajectories in the decision space, it suffers from the curse of dimensionality from the action space, making it infeasible for our problem setting (as we show later). To address these challenges, we present \textit{DG-MCTS} (domain-knowledge guided MCTS), which uses an action-generation framework that exploits domain knowledge and the underlying structure of the V2B optimization problem. This action-generation framework massively truncates the decision space. We draw from well-established heuristics and augment the set with randomly generated actions to ensure that the search tree is not limited to this truncated set. Collaborating with Nissan Advanced Technology Center - Silicon Valley (NATC-SV), we conduct extensive simulations on real-world data, showing that our approach outperforms existing approaches. To further ensure that our approach can scale to extremely large problem instances, we also propose a decentralized search algorithm that sacrifices performance (by a \textit{small} degree) to save computation time. In summary, our contributions are:
\begin{itemize}
    \item We present a decision-theoretic formulation for the V2B problem that models the interaction between electric vehicles and smart buildings as a stochastic control process.
    \item We show how MCTS, augmented with an action-generator module based on domain-specific heuristics and a randomized augmentation step, can compute near-optimal actions for the MDP.
    \item We also present a decentralized version of the algorithm that is significantly faster, albeit at the cost of a small deterioration in performance.
    \item We use real-world data in collaboration with NATC-SV and show that the proposed approach outperforms competitive baselines.
\end{itemize}

%% file: sections/03_related_work.tex

We highlight four major challenges of solving the V2B problem, namely: 1) the uncertainty due to EVs' arrival and departure times; 2) the dynamic energy costs, building loads under Time-of-Use (TOU) pricing; 3)  demand charges and long-term rewards; and 4) the heterogeneity of chargers;

\noindent\textbf{Inherent uncertainty:} Traditional optimization methods like Linear Programming (LP) and Mixed Integer Linear Programming (MILP) are widely used in V2B systems to minimize costs while meeting constraints \cite{kuang2017collaborative,tanguy2016optimization}. However, they assume precise knowledge of power usage and EV schedules, which is unrealistic in real-world applications.
To handle this uncertainty, research has explored stochastic optimization \cite{Sato2022OptimizationOE} and robust control methods \cite{Nguyen2023DistributionallyRM}. Probabilistic forecasting using models like Recurrent Neural Networks (RNN) can capture temporal dependencies \cite{Zhang2020DeepLearningBasedPF}. Reinforcement Learning (RL) has also been applied to learn adaptive charging strategies \cite{sultanuddin2023development, Abdullah2021ReinforcementLB}.
Despite these advances, managing long-term uncertainty remains challenging. Deep learning models struggle with long-term dependencies, and RL algorithms face sample inefficiency and convergence issues in complex environments \cite{Abdar2020ARO}. Limited exposure to rare events like extreme weather, sudden rate changes, or shifts in EV behavior can further reduce adaptability, leading to suboptimal solutions.

\noindent\textbf{Time-Of-Use (TOU) rates:} The variability of energy prices under TOU rates introduces additional complexity to the V2B optimization, requiring charging schedules that adapt to dynamic electricity costs, to minimize expenses \cite{Bitencourt2017OptimalEC}. The use of heuristics and machine learning algorithms, along with Deep Reinforcement Learning (DRL) have been useful in learning optimal charging policies responsive to fluctuating TOU rates \cite{Lpez2019DemandSideMU}. However, these methods face challenges due to difficulties in credit assignment and non-stationary energy usage patterns exhibited by both smart buildings and EV users~\cite{Lee2020ElectricVC}.

\noindent\textbf{Demand charges and long-term rewards:} Optimizing V2B is hard due to the mismatch between the prediction horizon and billing period. Existing work on demand charge prediction using model predictive control (MPC) is limited by the computational complexity of the long billing period. Thus, they resort to using shorter prediction horizons of a single day, resulting in suboptimal decisions, especially when the peak demand occurs outside the prediction horizon~\cite{risbeck2019economic,jianing2024}.

\noindent\textbf{Charger heterogeneity:} Smart buildings typically implement EV charging infrastructure incrementally, resulting in a diverse array of chargers with varying power ratings and directionality. This further complicates the action space for optimization. Some prior work addresses heterogeneity~\cite{ghafoori2023electricity}. However, they sacrifice long-term rewards by limiting planning to a single day or disregarding the presence of demand charges altogether.

\noindent\textbf{Role of Monte Carlo Tree Search}: Search-based algorithms like Monte Carlo Tree Search (MCTS) offer a promising alternative to handle inherent uncertainty in V2B systems. MCTS can manage stochastic environments by simulating numerous possible future scenarios, making it well-suited for planning under uncertainty \cite{pettet2020algorithmic, Bai2013BayesianMM}. Unlike traditional RL methods, MCTS can effectively plan over long horizons by building a search tree that considers future states and rewards \cite{pmlr-v235-schramm24a}. This capability allows it to handle long-term dependencies and adapt to sudden changes in EV behavior by continuously updating the search tree with new information \cite{Lucas2014FastEA}.



%% file: sections/02_problem.tex
We begin by describing our problem setting. Recall that in our setting, the EV owners follow a regular pattern of arrivals and departures with a predictable battery usage profile. While the building's energy usage is typically not known ahead of time, we assume access to a predictive model that uses historical data to predict the building's energy usage, which can be done very reliably~\cite{yin2024review,haophm2020}.

\subsection{Specifications}\label{ssec:problem_spec}

\noindent \textbf{Time Slots}: The total time under consideration (billing period) is divided into a finite set of discrete time slots. 
A time slot $t \in \mathcal{T}$ begins at $t^{\text {start}}$ and ends at $t^{\text{end}}$. The duration of each time slot is $\delta_t = t^{\text{end}} - t^{\text{start}}$, measured in hours.
Depending on the pricing policy of the power utility company, the hours of the day can be divided into \textit{peak} and \textit{off-peak} hours. The power utility usually computes the demand cost (explained later) during the peak hours $\mathcal{T}^{peak} \subset \mathcal{T}$, based on an aggregate of time slots of length $\tau$, i.e., $t^{\tau}_j = \{t_i, t_{i+1}, ..., t_{i+\tau}\}$ for the $j^{th}$ aggregate slot, and $i$ is an index to the corresponding time slot. The duration of each aggregate time slot is $\delta_a = \delta_t \cdot \tau$. The peak hours are then divided into $\Omega$ aggregated time slots $\{t^{\tau}_{1}, t^{\tau}_{2}, \cdots, t^{\tau}_{\Omega}\}$. We define all such sets of $t_i \in t^{\tau}_j$ using the notation $\mathcal{T}^{\tau}_j$. For example, if the peak hours are from 12 am to 1 am, each time slot is 5 minutes, and $\tau$ is 15 minutes, then the 1st aggregate time slot ($j = 0$), which is from 12 am to 12:15 am, is represented as $t^{\tau}_{0} = \{t_{0}, t_{1}, t_{2}\}$, then  $\mathcal{T}^{\tau}_0 = \{0, 1, 2\}$.\\
\noindent \textbf{Time-of-Use Price}: Power companies usually compute costs based on two components or charges---an energy charge and a demand charge. Some companies have variable energy charge rates, where a time-of-use pricing model is used. The energy charge (\$/kWh) can vary between the peak and off-peak hours and is denoted by the set $\mathcal{W}$ where $w^e_t$ is the time of use price for time slot $t \in \mathcal{T}$.\\
\noindent \textbf{Demand Charge}:
The other part of cost computation is the demand charge $w^d$ (\$/kW). This is a fee based on the highest rate of electricity usage during a specific time period within a billing period. Often only usage during peak hours is considered. \\
\noindent \textbf{Vehicles}: We denote the set of EVs by  $\mathcal{V}$. We assume that a vehicle $v \in \mathcal{V}$ has a battery capacity of $\left [e^v_{min}, e^v_{max} \right ]$, with its SoC at time $t$ denoted as $e^v_t$. Each vehicle is available to charge between their arrival and departure time slots $\left [ t^v_{start}, t^v_{end} \right ]$; where $t^v_{end}$ is unknown to the solver, due to our stochastic problem formulation. Instead, the EV user provides a window of departure, $\mathcal{T}^v_{end}$. A user arrives with a state of charge (SoC) of $e^v_{t^v_{start}}$ and requires a SoC of $e^v_{req}$ at departure, out of which only $e^v_{t^v_{start}}$ is known at the time of arrival. The SoC a user departs with may differ from their requested value, $e^v_{req}$, and is denoted as $e^v_{t^v_{end}}$. We later show how we compensate the users if their requirements are not met. Our EV partner's vehicles support bidirectional operations (charging and discharging), and the problem can be modified to serve only unidirectional (charging only) settings as well. The charging rate of an EV is represented as $c^v_t$. The power used to charge the EV over an aggregate time slot is represented as $c^v_{t^{\pi}_j} = \sum_{i=1}^{\tau} c^v_{t_i}$.\\
\noindent\textbf{Building's Power Consumption}: We denote the building's energy consumption using $\mathcal{B}^e = \{b^e_1,b^e_2,\cdots,b^e_{|\mathcal{T}|}\}$ in kWh. It is used to compute the energy charge by the power company.
The power draw during peak hours is represented by
$\mathcal{B}^p = \{ {b^p_0}, {b^p_{1}}, \cdots, {b^p_{\Omega}}\}$, at each aggregate time slot $t^{\tau}_j$, which covers $t_i, \cdots, t_{i+\tau}$ time slots, $b^p_j = 1/\tau \cdot \sum_{i=1}^{\tau}({b^e_i}/{\delta_t})$ and is used to compute the demand cost.\\
\noindent \textbf{Charging:} Chargers vary in \textbf{control mode} (controlled or uncontrolled), \textbf{directionality} (unidirectional or bidirectional), and \textbf{maximum rate} (e.g., 10 kWh, 20 kWh). Controlled chargers can start and stop charging as needed, while uncontrolled chargers charge continuously until full. Unidirectional chargers only supply power to vehicles, while bidirectional chargers also discharge, enabling V2B charging.
Each charger type is defined by a combination of these attributes (e.g., a \textbf{20 kWh controlled unidirectional charger}). The set of all charger types is represented as $\mathcal{K}$, with a total of $N$ chargers. Each type $k \in \mathcal{K}$ has $|k|$ chargers available, such that $\sum{|k|} = N$. Chargers follow a \textbf{linear charging profile}, uniformly charging an EV at each time step, which has minimal impact on optimization~\citep{sundstrom2010optimization}. We later compare this with a \textbf{piecewise charging curve} in an ablation study.
\\

\noindent \textbf{Charging Rate}: We define charging rate as the amount of power delivered to the EVs at each time slot, measured in kWh. We assume that the EVs can charge and discharge at the maximum rate supported by the charger, which ranges from $[q_{k}^{min}, q_{k}^{max}]\  \forall k \in \mathcal{K}$, where $q_{k}^{min}$ can be negative if the charger is bidirectional, denoting the discharging of a connected EV. A charger's efficiency is denoted by $\eta$, which is applicable to both charging and discharging. We also maintain an EV to charger type occupancy function $\zeta: \mathcal{V} \times \mathcal{T} \rightarrow \mathcal{C}$, where $\zeta^{v}(t) = k_i$, representing the connection of EV $v$ to a charger of type $k_i$  at time $t$. Once a charger is assigned to an EV, it cannot switch chargers until departure.\\
\noindent \textbf{Peak Power Use}: The peak power over an aggregate time slot $t^{\tau}_j$ is denoted by $\pi_j$. Recall that an aggregated slot $t^{\tau}_j$ consists of smaller slots $\{t_i, t_{i+1}, ..., t_{i+\tau}\}$; the peak power is computed by averaging the power draw (sum of power used by the building and the chargers) across each aggregate time slot $t^{\tau}_j$, $\pi_j = b^p_j + \sum_{i=1}^{N} c^v_{t^{\pi}_j}$. Thus, peak (maximum) power used during the peak hours is $P^{max} = \max(\pi_j)$.\\
\noindent \textbf{Demand Cost}:
Demand cost is levied by electricity suppliers based on a customer's peak rate of electricity consumption (power), typically measured in kilowatts (kW), over a specific period, usually a month. This charge is separate from the energy cost, which is based on the total amount of energy used (kWh). 
It is the product of the peak power in an aggregate time slot and the demand charge, and is computed as $w^d \cdot P^{max}$. 
It is difficult for models to optimize and plan for demand costs over longer billing periods due to the uncertainty and inflexibility of its assessment. Thus, demand costs need to be calculated non-myopically.\\
\noindent \textbf{Energy Cost}: The total energy used in time slot $s$ is the sum of the energy used in recharging the vehicles and meeting the building's energy requirement. We denote the energy cost for the time slot using $g_t  = w_t \cdot \sum_{v \in \mathcal{V}}(c^v_t + b^e_t)$.
By summing this across all slots over the billing period, we get the total energy usage cost. 
During time slots with high electricity prices, we can effectively reduce the overall bill while meeting the building's energy needs by discharging connected EVs. These EVs are then charged at a later time when the electricity price is lower.\\
\noindent \textbf{Missing SoC Cost}: The energy shortfall between required and actual SoC at departure for each EV is a key metric in the V2B problem. In our setting, since the departure of each EV is unknown, this metric is important to understand how well the user's requirements are met. We assign a monetary value to the missing energy, at the rate of  $w^s$, in \$/missing kWh.\\
\noindent \textbf{Total Bill}: The total bill over the billing period is the sum of demand cost and energy cost. The intuitive strategy is to minimize the peak power over the billing period, which reduces the demand cost and ensures that the energy cost during peak hours remains the minimum required to meet the constraints of charging the EVs to the required SoC level. It is represented as:
    \begin{equation}\label{eq:total_cost}
    \sum_{t\in\mathcal{T}} w^e_t \cdot (b^e_t + c^v_t) + w^d \cdot P^{max} + \sum_{v \in \mathcal{V}} w^s \cdot |e^v_{t^v_{end}} - e^v_{req}|        
    \end{equation}
\\
\noindent \textbf{Solution Space:} We define a solution with an assignment of chargers to cars and the corresponding charging/discharging schedule. Let $\mathcal{H}$ be the set of solutions, where for each charging assignment $(k, t)$, exactly one EV, $v \in \mathcal{V}$ is assigned, such that $\langle v, k, t \rangle \in \mathcal{H}$, as is the case in practical usage, where one EV is attached to only charger and not switched around. For each assignment $h_{v, k, t} \in \mathcal{H}$, we assign a charging rate $c^v_t \in C$. We consistently use the assumption that the EV remains at the charger for the entire duration of its stay. We assign cars on a \textit{first-come first-serve basis}, where we prioritize assigning cars based on their time of arrival and chargers by their rate of charging, directionality, and controllability. For example, one priority list could be assigning cars to 20kWh bidirectional, 20kWh unidirectional (all controlled), and then the uncontrolled chargers. We can accommodate only as many cars as available chargers. Excess cars that arrive when there are no vacant chargers, will not charge for the remainder of their stay.

%% file: sections/02b_mdp.tex
\subsection{Markov Decision Process (MDP)}

We model the V2B problem as a Markov Decision Process (MDP), building on prior work by~\citet{shi2011real}. An MDP is characterized by a 4-tuple $\{\mathcal{S}, \mathcal{A}, \mathcal{P}, \rho\}$, where $\mathcal{S}$ is a set of states, $\mathcal{A}$ is a set of possible actions, $\mathcal{P}$ is a state-action transition model, and $\rho$ is a reward function which captures the agent's utility (or cost)~\citep{Kochenderfer_decision}.  \\
\noindent\textbf{Decision Epoch}: A decision epoch occurs at every discrete decision-making event, $t \in \mathcal{T}$. Between events, the environment moves in continuous time where chargers charge or discharge EVs.
At each decision epoch, the decision-maker takes an action that moves the state from a pre-decision state to a post-decision state and is then given an immediate reward.\\
\noindent\textbf{State}: We denote the set of states as $\mathcal{S} \in \{\{b^e_t\}, \hat{P}^{max}, \{q_k^{min}, q_k^{max}\}$  $ \forall k \in \mathcal{K}, \{e^v_t, t^v_{start},  \hat{t^v}_{end}, e^v_{min}, e^v_{max}\} \forall v \in \mathcal{V}\}$ and a state at time slot $t$ is represented as $s_t$ at pre-decision time. It includes the current building load ($b^e_t$), all charger rates ($q_k^{min}, q_k^{max}$), EV details ($e^v_t, t^v_{start}, \hat{t^v}_{end}, e^v_{min}, e^v_{max}$), where $\hat{t^v}_{end}$ is the estimated departure time from user's departure time window $\mathcal{T}^v_{end}$. We also include an estimate of the peak power $\hat{P}^{max}$ to assist in making better decisions.\\
\noindent\textbf{Actions}: We denote the set of all feasible actions at time slot $t$ by $\mathcal{A}_t$. An action in $\mathcal{A}_t$ corresponds to a combination of charging rates for all chargers $\mathcal{K}$. Charging rates are limited by the total power consumed by the building, including the power provided for charging the vehicles, i.e., the EVs cannot be discharged more than the building's current power usage. They are limited to discrete values between the maximum and minimum allowed charging rates.\\
\noindent\textbf{State Transitions}: At time slot $t$, the decision-maker can take an action that results in a transition from a state $s$ to $s'$. In this transition, the system evolves through many different stochastic processes. First, EVs can arrive or depart out of their regular arrival and departure times, and are governed by some duration of stay distribution. Second, electricity prices may change with minimal warning, as may happen during \textit{emergency load reduction programs} (ELRP) events wherein consumers receive financial incentives for reducing their energy consumption.
Finally, the building's power draw may change depending on the time of day, month of year, or even weather. While these follow a predicted distribution, they still introduce stochasticity in state transitions. 
We omit detailed discussion of the mathematical model and expressions for temporal transitions and state transition probabilities, as our framework relies solely on a generative world model rather than explicit estimates of these transitions.

\noindent\textbf{Rewards}: Rewards in an MDP often have two components: a lump sum immediate reward for taking actions and a continuous time reward as the process evolves, and are highly domain-dependent. For charger optimization, we are concerned about minimizing the overall energy usage cost while meeting a certain quality of service for users. We denote the reward function by $\rho(s, a)$, for taking action $a$ at state $s$. The reward is both intermediate $r^i_t$ and episodic $r^e_t$. We use episodic rewards during the rollouts. The intermediate reward is $r^i_t = g_t + w^s \cdot \sum_{v \in \mathcal{V}} \left |e^v_{t^v_{end}} - e^v_{t^v} \right |$ if $t = t^v_{end}$, else, $r^i_t = g_t$. The episodic reward $r^e_t$ is calculated according to equation~\eqref{eq:total_cost}, with a minor modification to the demand cost calculation which uses an estimate of the peak power, $\hat{P}^{max}$. It is as follows:
{\small
\begin{equation}    
    r^e_t = \sum_{t\in\mathcal{T}} w^e_t \cdot (b^e_t + c^v_t) + w^d \cdot \hat{P}^{max} + \sum_{v \in \mathcal{V}} w^s \cdot |e^v_{t^v_{end}} - e^v_{req}|   
\end{equation}
}

%% file: sections/04b_online_approach.tex

We propose an online approach for managing charging controls. First, we use a heuristic EV assignment policy, such as \textit{first-come, first-serve} and bidirectional-first policies, to assign EVs to chargers while considering fairness and the peak shaving capability of bidirectional chargers. To address uncertainty in EV arrival and departure, we estimate future system states by sampling data and utilizing an offline solver to derive optimal actions and establish upper bounds on performance metrics (described in the Appendix, Section~\ref{sec:offline_app}.

To adapt to environmental uncertainty, we employ an online MCTS search for dynamic and robust decision-making. Recognizing the challenge of MCTS runtime in real-time scenarios, we incorporate domain-knowledge guidance (DG-MCTS) to shrink and adjust the action space using heuristic actions and demand charge predictions, improving exploration efficiency. 
Our online DG-MCTS solver constructs a forward-looking search tree using episodic data provided by our EV operating partner, \nissan{}. To reduce computational complexity, we decompose long billing periods (monthly) into shorter, manageable daily planning horizons. Algorithm~\ref{algo:MCTS} provides an overview of the workflow of our DG-MCTS approach. In the following section, we detail each component of the approach, including EV assignment, future state estimation, and the integration of domain-knowledge-guided exploration into MCTS.


\subsection{Episode Sampling and Handling Uncertainty}
\noindent \textbf{Episode Samples:} We modeled a Poisson distribution using real-world data from Nissan Advanced Technology Center - Silicon Valley (NATC-SV) in Santa Clara, California, USA.

%
%
%
\noindent \textbf{Car to Charger Assignment:}
We assign EVs to chargers based on a \textit{first-come-first-serve} policy based on the arrival time. If multiple cars arrive in the same time slot, we break ties by ordering them according to $e^v_{req}$. EVs are assigned to chargers based on the charger's directionality, maximum rate, and controllability. \\
%
\noindent \textbf{Uncertainty in Departure Times:}
Given that each user only provides a departure window $\mathcal{T}^v_{end}$, our goal is to estimate the departure time within this interval. Since $\mathcal{T}^v_{end}$ represents a range of possible departure times rather than a fixed point, we model this uncertainty by treating the departure time as a random variable within $\mathcal{T}^v_{end}$. We draw samples from a uniform distribution over the interval $\mathcal{T}^v_{end}$ to approximate a representative departure time.

Let $\hat{t}^v_{end}$ denote the estimated departure time for user $v$ as a realization of a uniformly distributed random variable over $\mathcal{T}^v_{end}$. We sample $\hat{t}^v_{end} \sim \mathcal{U}(\min(\mathcal{T}^v_{end}), \max(\mathcal{T}^v_{end}))$ where $\min(\mathcal{T}^v_{end})$ and $\max(\mathcal{T}^v_{end})$ are the lower and upper bounds of the departure window $\mathcal{T}^v_{end}$, respectively. Sampling in this manner allows us to select a feasible departure time that is unbiased with respect to any specific point within the interval, ensuring a fair estimation across the entire window. This serves as a proxy for user departure behavior.

\noindent \textbf{Estimating Peak Power Threshold}: 
For a fixed EV arrival and departure trajectory, we determine optimal charging decisions by solving a MILP, minimizing costs while meeting charging requirements. This process also estimates the monthly peak power threshold. We solve sampled episodes, distinct from evaluation data, and compute peak power from optimal actions. The 99\% confidence level of peak power across samples provides a robust monthly estimate. This confidence-based threshold integrates demand charges into the sampling process, aligning with optimal charging outcomes from the MILP solutions.

\subsection{Handling Exponential Action Space}
\label{ssec:actionspace}
%
A feasible action in our problem corresponds to a set of charger rates for all chargers, given that chargers can be heterogeneous with the possibility of turning off,  charging, and discharging a connected EV. Additionally, the sum of charger rates must be constrained to the current building load at that time. As a result, for a building with a large number of chargers with varying configurations, the possible actions for a given state are combinatorially large, $N^{p}$ if we consider $p$ discrete actions for each charger. Such an action space is infeasible to explore in an online setting. To address this challenge, we introduce a heuristic that enables us to identify promising actions from the set of feasible actions.\\
%
%
\noindent\textbf{Least Laxity First heuristic:} One of our goals is to charge EVs such that they leave with their desired SoC level. To guide the charging decisions, we utilize the least laxity first (LLF) heuristic~\cite{xu2016dynamic}. This heuristic prioritizes EVs with the least remaining time until departure, ensuring that vehicles close to their departure window receive priority in charging allocation. 
At each decision epoch, we compute the available power gap at the current state $s$ and time $t$ as $\text{power gap} = \hat{P}^{\text{max}} - b^p_t$, where $\hat{P}^{\text{max}}$ is the estimated peak power threshold and $b^p_t$ is the current building load. The trickle charging rate for each EV $v$ is defined as $\text{trickle rate} = (e^v_{\text{req}} - e^v_t) / (\hat{t}^v_{\text{end}} - t)$, where $e^v_{\text{req}}$ is the required energy, $e^v_t$ is the current energy, and $\hat{t}^v_{\text{end}}$ is the estimated departure time.
We then calculate the sum of trickle rates for all EVs at the current time slot. If this sum is less than the available power gap, we have the capacity for overcharging. In such cases, we set each EV’s charging rate to its trickle rate and assign additional charging to EVs with bidirectional chargers, following a reverse order of laxity (least laxity first) until the power gap is fully utilized. If trickle rates exceed the power gap, we discharge EVs on bidirectional chargers, prioritizing those with the most time before departure, to fill the gap before resuming trickle charging. 

%
We improve the promising actions by introducing intuitive actions based on two parameters, the power gap and the bounds of feasible actions. If the available power gap is \textit{positive}, indicating surplus capacity, we add discrete incremental charging actions to explore overcharging options. Otherwise, if the power gap is negative (where the building load exceeds the estimated peak power), additional discharging actions are included to mitigate the excess demand; we refer to these additional actions as $\beta$. Finally, we add an action between the minimum chosen action and the minimum feasible charging rate, $q_k^{\text{min}}$, ensuring that it does not go below $e^v_{\text{min}}$, the minimum state of charge required by the EV. Similarly, we add another action between the maximum chosen action and the maximum feasible charging rate, $q_k^{\text{max}}$, ensuring that it does not exceed $e^v_{\text{max}}$, the EV’s maximum allowable state of charge; we refer to this change in action-space as ``offset''.


This LLF-guided action serves a dual purpose: it enhances traceability by narrowing the search space and accelerating decision-making. This keeps the search process computationally fast. By prioritizing actions based on current states, including the urgency of each EV, the LLF heuristic enables us to adapt charging rates quickly, focusing exploration on the most relevant actions.
%

\input{algorithms/centralized_MCTS}

\noindent\textbf{Temporal Decomposition:}
We decompose the long billing period (typically a month) into shorter, computationally manageable planning horizons, such as a day. However, this decomposition poses a significant challenge: the demand charge can only be accurately calculated at the end of the full billing period, i.e., over the longer planning horizon. To address this, we leverage two essential properties of the V2B Markov Decision Process (MDP).

First, note that the arrival and departure times of EVs are independent of the agents' charging actions—EVs arrive and depart from the building regardless of how existing vehicles are charged. This independence allows us to \textit{pre-sample} trajectories of EV arrivals and departures using a generative model based on historical data, without conditioning the sampling on the actions taken within the search tree.

Second, we leverage the estimated peak power threshold, as described earlier. 
Equipped with this peak power estimate, we introduce the core concept enabling temporal decomposition: divide the full planning horizon into smaller periods, with the added constraint that the total power consumed in each period remains below the estimated demand charge for a fixed sampled trajectory.

At the beginning of each day (the smaller decomposed planning horizon), we sample multiple trajectories of EV arrivals and departures from the generative model. For each trajectory, the trained model $f$ estimates the demand charge over the longer planning horizon (e.g., a month). This estimated charge is then integrated and used as part of the episodic reward, within the search tree, which operates on the shorter daily planning horizon.


\subsection{Monte Carlo Tree Search (MCTS) Evaluation}
Offline approaches to solving the V2B problem such as MILP, fail to consider the stochasticisity present in the real world. Instead, they rely on complete knowledge of the system to optimally select the best actions. This motivates us to use MCTS, an anytime algorithm that has been widely used in game-playing scenarios~\citep{silver2018general}. 

MCTS models planning as a tree with states as nodes and actions as edges. It explores the tree asymmetrically, favoring promising actions with the Upper Confidence bounds applied to Trees (UCT) algorithm \cite{kocsis2006bandit}, balancing exploitation and exploration. Node values are estimated through \textit{rollouts} using a simple default policy, often random action selection. As the search progresses, node value estimates improve. This approach enables efficient exploration of large action spaces. MCTS requires a generative environment model, a tree policy for navigation, and a default policy for value estimation.

We use the collected historical data to sample new episodes as the tree is built into the future. We use the standard Upper Confidence bound for Trees (UCT)~\cite{UCT} to navigate the search tree and decide which nodes to expand. When expanding a node we sample promising actions for the given state. When working outside the MCTS tree to estimate the value of an action during rollout, we rely on a default policy. This policy is simulated up to a time horizon and the utility is propagated up the tree. Our default policy is a trickle charging rate policy -- which charges each car with the required energy to meet the required SoC by the estimated departure time. \\
\noindent\textbf{Root Parallelization:} Given that EV arrivals and departures and building power draw, even when following a known distribution, are highly uncertain in time, sampling one episode may not represent actual future EV behavior. We handle this using \textit{root parallelization}, which involves sampling many episodes, and instantiating a new MCTS tree for each with their EV arrivals/departures and building power draw as the root node. Each tree is explored in parallel, and after execution, the score for each of the actions from the common root node is averaged across the trees. The action with the highest average score across all trees is then the selected action.


%% file: algorithms/centralized_MCTS.tex
\begin{algorithm}[htbp]
\raggedright
\small
\caption{Domain-knowledge Guided MCTS (DG-MCTS)}
\label{algo:MCTS}
\KwIn{Current State $\mathcal{S}_t$, iterations $\mathcal{I}$, exploration range $\beta$}
\KwOut{Best action $[\mathcal{A}^*_k \text{ for } k \in K]$, }

$D \gets \text{EstimateDemandCharge}(\mathcal{S}_t)$ \tcp{Estimate demand charge}
$[\mathcal{A}'_k \text{ for } k \in K] \gets \text{LLFHeuristic}(\mathcal{S}_t, D)$ \tcp{Get heuristic actions}
\tcp{Action Pruning}
\If{$BuildingLoad < D $ }
{\ForEach(\tcp*[h]{Encourage Charging}){$k \in K$}{ 
    $Space(\mathcal{A}_k) \gets \text{GetNeighbors}
    ([\mathcal{A}'_k - \beta, \mathcal{A}'_k + \beta + \text{offset}] \cup \text{NearBoundaryActions})$ 
}}
\Else{
\ForEach(\tcp*[h]{Encourage Discharging}){$k \in K$}{
    $Space(\mathcal{A}_k) \gets \text{GetNeighbors}([\mathcal{A}'_k - \beta-\text{offset}, \mathcal{A}'_k + \beta] \cup \text{NearBoundaryActions})$ 
}
} 
$ActionSpace \gets [Space(\mathcal{A}_K). Clip[q_k^{min}, q_k^{max}] \text{ for } k \in K]$ 

\For(\tcp*[h]{Establish tree}){$n \gets 1$ to $\mathcal{I}$}{
    $Sample \gets \text{GenerateSample}( \text{EVArrTime}, \text{EVDepTime}, \text{BuildingLoad})$ \tcp{Move to next state, apply pruned action}

    $\mathcal{S}_{t+1}' \gets \text{SelectFrom}(\mathcal{S}_t, ActionSpace, Sample)$ 
    \tcp{Move to next state, apply pruned action}

    $\mathcal{S}_{t+2}' \gets \text{Expand}(\mathcal{S}_{t+1}',  Sample)$ \tcp{Add new child node}
    $v \gets \text{Simulate}(\mathcal{S}_{t+2}', Sample, \text{TrickleRate})$ \tcp{Rollout}
    $\text{Backpropagate}(\mathcal{S}_{t+2}', v)$ \tcp{Update tree stats}
}

$\mathcal{A}^* \gets \text{BestAction}(\mathcal{S}_t)$ \tcp{Select action with highest value} 
\end{algorithm}

%% file: sections/04c_MCTS.tex

\subsubsection{Centralized MCTS}

In a centralized multi-agent approach, a single search tree represents the combined decision-making space of all agents in the V2B system. Rather than each agent operating independently, this unified tree integrates the actions of all agents, allowing for joint optimization. Such an approach is advantageous in V2B settings, where decisions made by individual EVs impact overall building load and demand charges, requiring coordination to minimize costs. However, the centralized MCTS also faces computational challenges due to the high dimensionality of the action space and the extended planning horizon, as it must simultaneously consider all agents’ actions at each decision point. We show the process in Algorithm~\ref{algo:MCTS}.

Centralized MCTS explicitly considers the interactions among agents at each step, which is particularly advantageous in scenarios where joint coordination is necessary. For example, centralized MCTS can directly incorporate the influence of one EV's charging actions on the demand charge for all EVs and the building. However, centralized MCTS must address two primary challenges: \textbf{(1)} efficiently navigating the large action space within a single tree structure and \textbf{(2)} managing the long temporal horizon associated with the V2B optimization problem. We tackle these challenges by leveraging domain-specific heuristics, which streamline the exploration of the action space and introduce strategic planning over shorter horizons within the MCTS framework, and name it Domain-knowledge Guided MCTS (DG-MCTS).

\subsubsection{Decentralized MCTS}
While DG-MCTS can address large state spaces, it still struggles with high-dimensional action spaces and long time horizons. As the number of cars grows, a single MCTS tree may become too large. To manage this, we use a decentralized approach to multi-agent MCTS (dMCTS) ~\cite{claes2017decentralised,pettet2020algorithmic}.

In dMCTS, each agent builds its own search tree to find a near-optimal action. This reduces the search space at each agent and helps the system scale. However, two challenges arise: \textbf{(1)} Each agent must consider the actions of other agents during exploration. If we rely on a proxy that predicts other agents’ actions ~\cite{pettet2020algorithmic}, we risk incurring high demand costs, \textbf{(2)} Decentralizing the action space does not fix the long horizon problem.

We address these challenges with a heuristic that leverages the structure of the Vehicle-to-Building (V2B) problem:

\noindent \textbf{Criticality Score}:  
We order agents by how urgent their charging needs are. We assume the total energy draw is limited, so we prioritize agents that must meet higher charge requirements in less time. For each connected EV $v$, we compute a criticality score using Least Laxity First~\cite{xu2016dynamic}, $c_{score} = ({t^v_{end} - t}) - (e^v_{req} - e^v_{t})/q^{max}_{\zeta^v(t)}$
where $t^v_{\text{end}}$ is the time slot when $v$ leaves, $t$ is the current time slot, $e^v_{\text{req}}$ is the required state of charge (SOC), $e^v_t$ is the current SOC, and $\zeta^v(t)$ is the charger connected to $v$. An agent with a smaller laxity (time left after reaching its required charge) has higher priority.

\noindent \textbf{Procedure}:  
Each charger uses dMCTS to make decisions without a centralized controller. This setup lowers computational cost because each agent focuses on its own action space, yet still accounts for the choices of others. We detail the steps for action selection, rollout, and backpropagation in Algorithm~\ref{algo:dmcts}. Algorithm~\ref{algo:sort_cars} shows how we sort cars by their criticality scores. This ordering ensures that cars with urgent needs get priority in the shared energy allocation, even though each agent runs its own local search.

\input{algorithms/decentralized_MCTS}
\input{algorithms/Sort_Cars}

%% file: algorithms/decentralized_MCTS.tex
\begin{algorithm}[htbp]
\raggedright
\small
\caption{Decentralized Monte Carlo Tree Search (dMCTS)}
\label{algo:dmcts}
\setcounter{AlgoLine}{0}
\KwIn{Current State $\mathcal{S}_t$, iteration number $\mathcal{N}$, exploration range $\beta$}
\KwOut{Best actions $[\mathcal{A}^*_k \text{ for } k \in K]$ for all chargers}
$D \gets \text{EstimateDemandCharge}(\mathcal{S}_t)$ \tcp{Estimate demand charge}

\ForEach{$k \in K$ sorted by least laxity first }{
    $\mathcal{A}'_k \gets \text{LLFPolicy}(\mathcal{S}^k_t, D)$ \tcp{Get heuristic action} 
        $Space(\mathcal{A}_k) \gets \text{GetNeighbors}([- \beta + \mathcal{A}'_k + \text{Noise}, \beta + \mathcal{A}'_k + \text{Noise}])$ 
    \For(\tcp*[h]{Establish tree for each charger}){$n \gets 1$ to $\mathcal{N}$}{
    
    $Sample \gets \text{GenerateSample}( \text{EVArrivalTime}, \text{EVDepartureTime}, \text{BuildLoad})$ 
       
        $\mathcal{S}^k_t\gets \mathcal{S}_t$ 
         
        $\mathcal{S}^k_{t+1} \gets \text{SelectFrom}(\mathcal{S}^k_t, Space(\mathcal{A}_k), Sample)$
        \tcp{Move to next state, apply pruned action}

        $\mathcal{S}^k_{t+2} \gets \text{Expand}(\mathcal{S}^k_{t+1}, Sample)$ \tcp{Add new child nod}
        $v \gets \text{Simulate}(\mathcal{S}^k_{t+2}, Sample, LLFPolicy)$ \tcp{Rollout}
        $\text{Backpropagate}(\mathcal{S}^k_{t+2}, v^k)$ \tcp{Update tree state for $k$}
    }

    $\mathcal{A}^*_k \gets \text{BestAction}(\mathcal{S}_t)$ \tcp{Select the best action for $k$}

    \text{$\mathcal{S}_t\gets \text{Update}(\mathcal{S}_t, \mathcal{A}^*_k)$} \tcp{Update state for the next charger}

}
\Return $[\mathcal{A}^*_k \text{ for } k \in K]$
\end{algorithm}

%% file: algorithms/Sort_Cars.tex
\begin{algorithm}[htbp]
\raggedright
\small
\caption{Sort Cars by Criticality}
    \setcounter{AlgoLine}{0}
\label{algo:sort_cars}
\KwIn{Set of cars $\mathcal{V}$}
\KwOut{Sorted list of cars $\mathcal{V}'$ based on criticality}
$\mathcal{V}^{critical} \gets \emptyset$ \tcp{Initialize list of critical scores}

\ForEach{$v \in \mathcal{V}$}{
    $c_v \gets \text{ComputeCriticality}(v)$ \tcp{Compute critical score}
    $\mathcal{V}^{critical} \gets \mathcal{V}^{critical} \cup \{(v, c_v)\}$ \tcp{Append car $v$ and its criticality score}
}

$\mathcal{V}' \gets \text{Sort}(\mathcal{V}^{critical}, \text{by } c_v \text{ descending})$ \tcp{Sort cars by criticality score}

\Return $\mathcal{V}'$ \tcp{Return sorted list of cars}
\end{algorithm}

%% file: sections/05_experiment.tex
\subsection{Setting}

\noindent \textbf{Data Collection}
To assess the effectiveness of our proposed method, we utilize data from the NATC-SV research laboratory. Optimization is limited to weekdays, as few employees work on weekends, and demand charges typically exclude them. Silicon Valley Power does not count Sundays in demand charge calculations. Our setup includes 10 unidirectional (20 kWh) chargers and 5 bidirectional chargers supporting ±20 kWh (charging at 20 kWh and discharging at 20 kWh). All EVs have bidirectional charging capability.

\begin{figure}[htbp]
    \centering
    \includegraphics[width=0.95\textwidth]{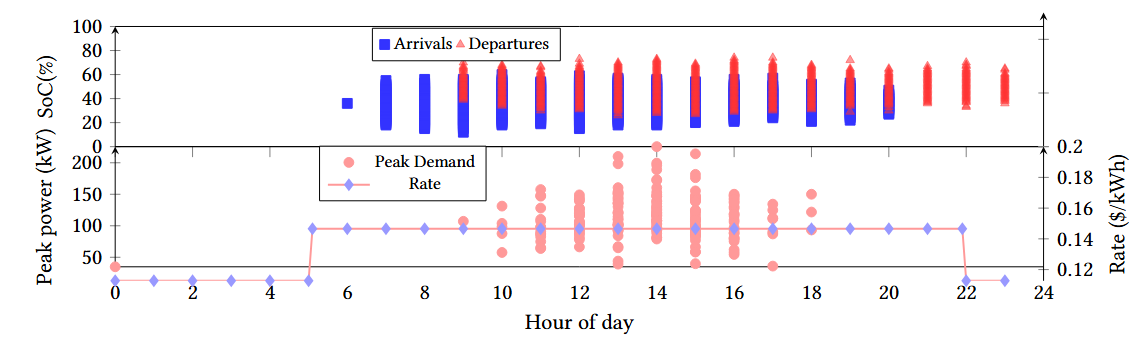}
    \caption{(Top) EV arrival and departure hours vs. arrival and required SoC over 8 months. (Bottom) Peak building power draw vs. time of day and TOU rates.}
    \label{fig:car_distributions}
\end{figure}

\label{text:q2}
We collected real-world data from NATC-SV in Santa Clara, California, USA, covering building power consumption, EV charger usage, and EV telemetry over a nine-month period from May 2023 to January 2024. \blue{Building load fluctuations are reliably predicted using standard models \cite{yin2024review}}, while EV behavior (arrivals, departures, and SoC requirements) is modeled using a Poisson distribution based on historical data. \blue{This approach captures variations in EV arrival rates and SoC demands, using hourly mean values to reflect realistic usage patterns.}  
The number of bidirectional EVs arriving on weekdays fluctuates daily, introducing inherent uncertainty. \blue{Vehicle arrivals and departures are user-specific behaviors, largely governed by work hours, and independent of building load or charging actions. In collaboration with the building operator, we confirmed that arrival and departure patterns are shaped by user decisions rather than system-level controls.}  
\blue{This independence allows arrival and departure times to be pre-sampled without impacting optimization. Additionally, current charging decisions do not affect future user arrival or departure times, and the optimization process must account for unpredictable user behavior while developing efficient charging plans within the existing limitations..}
Figure~\ref{fig:car_distributions} displays the arrival and departure times and the SoC requirement, along with the distribution of peak power demand and corresponding hours. We sampled $110$ monthly billing episodes from May 2023 to December 2023 to construct a representative dataset.\\
\blue{Electricity prices follow Silicon Valley Power rates in Santa Clara: peak hours are $\$0.147$/kWh (6 am–10 pm, except Sunday) and off-peak hours at $\$0.113$/kWh, with a $\$9.62$/kW demand charge during the peak hours.}

\noindent \textbf{Estimated Peak Power.} To improve action effectiveness, we account for varying weekday conditions by incorporating a monthly peak power estimate for each episode, based on optimal action sequences generated by the MILP solver. We use the lower bound of the 99\% confidence interval from the MILP data as a conservative estimate of the demand charge. This input feature is further optimized during RL training.\\

\input{tables/hyper}
\noindent\textbf{Hyperparameter Tuning.} To optimize the performance of our DG-MCTS framework, we utilized a state-of-the-art hyperparameter optimization library, Optuna, which employs efficient sampling strategies to explore the hyperparameter space. It supports algorithms such as Tree-structured Parzen Estimators (TPE) to adaptively sample promising regions of the hyperparameter space while discarding less effective configurations. Each trial in Optuna represents a unique combination of hyperparameters, which is evaluated based on the objective function, and results are stored in a study database for further analysis. Optuna's objective was to minimize a custom-defined score, calculated according to equation~\eqref{eq:total_cost}. The hyperparameter search space included key parameters: number of iterations, maximum depth, penalties for unmet SoC requirements and power gap exceedances, and rewards for meeting specific SoC targets. Additionally, we explored regularization parameters like $C$ and reward discounting factor $\gamma$, and a tolerance parameter for estimated peak power ($\epsilon$). Optuna's trial-based approach generated a hyperparameter importance graph as shown in Appendix in Figure~\ref{fig:optuna}, revealing the most influential factors affecting performance, and guiding subsequent model refinement. We set the parameters as shown in Table~\ref{tab:hyper_conf}.



\noindent \textbf{Hardware Used.} All the experiments were performed and timed on a 32-core 4.5 GHz machine with 128 GB of RAM.

\input{tables/main_table}
\input{tables/missing_soc}
\input{tables/cars_missing}
\input{tables/peak_shaving}

\noindent \textbf{Baseline Approaches.} We evaluate the performance of our online approach by comparing it against various methods including real-world charging procedures, several smart heuristic approaches, and a reinforcement learning-based policy. We provide a brief description of the baselines here.

\begin{itemize}[leftmargin=*]
    \item {\bf MaxCharge}: This approach simulates current real-world charging, where all connected EVs at the fastest rate to $e^v_{max}$.
    \item {\bf ReqCharge}: Similar to \textit{MaxCharge}, however only charges all connected EVs as quickly as possible to $e^v_{req}$.
    \item {\bf Least Laxity First (LLF)}: Uses the same heuristic policy used in Section~\ref{ssec:actionspace}. It charges beyond $e^v_{req}$ if possible and then leverages excess energy to reduce peak power demand. Otherwise, it uses trickle charging to charge EVs to $e^v_{req}$.
    \item{\bf Early Deadline First (EDF)}: EDF prioritizes EVs with the nearest departure, following the Early Deadline First scheduling approach~\cite{stankovic_EDF}. Like LLF, it may charge in excess if high building demand is expected and discharge before departure. Otherwise, it uses trickle charging to reach $e^v_{req}$.
    \item{\bf Reinforcement Learning (RL)}: We use the Deep Deterministic Policy Gradient (DDPG) algorithm to manage charging actions. The state includes time, building load, charging status, EV SoC, and expected departure times. The reward function combines demand cost, energy cost, and SoC deviation penalties.
    The DDPG model employs a two-layer Multi-Layer Perceptron (MLP) for actor and critic networks, each with 96 neurons. Action masking prevents charging when no EV is connected, ensures SoC targets are met before departure, and discharges excess energy. Policy guidance integrates MILP-generated optimal actions into training to improve performance. A separate model is trained for each month using 60 simulated samples. Table~\ref{tab:hyperparameters} details the hyperparameters.

\end{itemize}

\subsection{Results}

We evaluate all approaches using four metrics: 
\begin{enumerate}
    \item \textbf{Total Cost}: The sum of electricity cost, demand charge, and missing SoC cost for the billing period, according to Eq.~(\ref{eq:total_cost}).  
    \item \textbf{Missing SoC}: This is the energy shortfall of departing cars relative to their required SoC, and results in a penalty of 20 cents per kWh, 42\% higher than the grid’s maximum energy cost. This discourages the building from relying on bidirectional EVs to reduce grid energy purchases.
    \item \textbf{Cars Under Required SoC}: The number of cars with missing SoC shows if the model meets all required SoC targets or prioritizes some cars over others.
    \item \textbf{Peak Shaving}: It is the difference in demand charge between (i) the building's power usage (without any charging) and (ii) by adding charging the EVs under the respective policies. Positive values indicate the policy reduced the demand charge by controlling the charging actions.
\end{enumerate}

We evaluated our online approach using data from May to December 2023. Ten episodes were randomly selected as the test set, while the remaining 100 were used as exploration samples in root parallelization. Policies lacked actual vehicle departure times, relying instead on the departure time windows.

Table~\ref{table:monthly_policy_costs} compares monthly bills across eight months for different policies. Domain-knowledge-guided MCTS (DG-MCTS) outperformed other heuristics in six months, showing greater cost-effectiveness than real-world baselines MaxCharge and ReqCharge, and smart heuristics like LLF and EDF.

DG-MCTS effectively reduced monthly costs by optimizing peak shaving while maintaining missed SoC values comparable to the heuristic methods. It required an average of 23.75 seconds per decision, whereas dMCTS was faster at 15.38 seconds.

RL-based policies outperformed DG-MCTS in two of eight months, but many EV SoC requirements remained unmet (Table~\ref{tab:missing_soc}). Table~\ref{tab:missing_soc_by_policy} shows RL fails to meet SoC targets for an average of 207 cars due to action selection uncertainty. While our approach results in more missed SoC in some episodes, it performs similarly to smart heuristics in EVs falling short of required SoC.

Results indicate that our approach outperformed heuristics by optimizing actions under uncertainty. While smart heuristics improve on MaxCharge and ReqCharge, they cannot anticipate future events, limiting performance. Even with MaxCharge, some cars miss their SoC targets if their stay is too short.
 

\input{tables/ablation}
\input{tables/high_dep_var}
\input{tables/25cars}
\input{tables/future_sweep}

\label{text:q4}
\noindent\textbf{Robustness to Uncertainty.} Finally, we show that our approach outperforms all heuristics when we further increase the degree of uncertainty by: \textbf{1)} Increasing the potential departure window as shown in Table~\ref{tab:high_variance} and \textbf{2)} Handling an unexpected increase in the number of daily EV arrivals, by increasing the number of daily cars to be around 25 per day, as shown in Table~\ref{tab:25cars}. DG-MCTS tackles changes in the environment better than the rest of the policies, taking 48.97 seconds per decision while dMCTS takes an average of 25.46 seconds per decision. \blue{This highlights the capability of dMCTS to scale more effectively than DG-MCTS in terms of computation time. Table~~\ref{tab:25cars} also shows that dMCTS outperforms the rest of the baseline methods during this scale up}.












\subsection{Ablation Study}

The impact of key techniques in our approach is evaluated through ablation. We select August 2023, the month with the highest building peak, and assess its effect on the total bill. The ablations tested are: \textbf{1)} \textbf{\text{MCTS}/P}, which replaces predicted peak demand with only the current peak demand. \textbf{2)} \textbf{\text{MCTS}/H}, which removes heuristics for narrowing the action space. \textbf{3)} \textbf{\text{MCTS}/C}, which uses a piecewise-linear battery profile instead of a simplified linear model. Table~\ref{table:ablation} presents the impact of each feature.

\noindent\textbf{Peak Demand Prediction}: We examine \textbf{\text{MCTS}/P}, where peak demand prediction is removed in lieu of using the current power demand as the threshold, and the performance drops drastically. This shows the importance of having an accurate model for predicting peak demand for any length of the planning horizon.\\
\noindent\textbf{Heuristics for Action Pruning}: The \textbf{\text{MCTS}/H} approach removes the use of action pruning via heuristics, resulting in decreased performance. This shows the dependency of MCTS on the quality of the generated trees. Removing action pruning increases the action space, making it difficult to choose ``good'' actions. Increasing iterations can improve solutions but may increase computation time.\\
\noindent\textbf{Piecewise Linear Charging Profiles}: \textbf{\text{MCTS}/C} with piecewise linear charging profiles offers similar solutions but adds complexity. Most EV manufacturers do not disclose charging profiles, complicating SoC curve analysis. While piecewise linear profiles may improve accuracy, their absence does not affect the approach's validity, as noted by \citet{sundstrom2010optimization}.

\label{text:q1}
\blue{
\subsection{Sensitivity Analysis}
\label{ssec:sensitivity}
This section explores the effects of changing the accuracy of the EV behavior and building load predictions by perturbing the MCTS exploration samples. 
We evaluate three scenarios with modified exploration samples: (1) more EVs (ME), (2) fewer EVs (FE), and (3) higher building load fluctuations (BLF). The policy’s performance is tested on the 10 unchanged August test episodes from Table~\ref{table:monthly_policy_costs}. The results, in the right half of Table~\ref{tab:ablation_study}, assess the policy’s response to increased uncertainty and its impact on total cost.
}
\\
\blue{
When DG-MCTS is combined with exploration samples that include more EVs, denoted as MCTS(ME), the exploration samples contain 25\% more cars on average than the test episodes. In contrast, when using fewer EVs, represented as MCTS(FE), the exploration samples have 25\% fewer cars on average. While both scenarios result in a higher total bill than DG-MCTS, they still achieve a lower overall cost compared to the best domain-specific heuristic, LLF. 
In the MCTS(BLF) scenario, DG-MCTS is tested with exploration samples that have greater building load fluctuations, introduced by adding a fixed perturbation of ±10\% to the predicted load. This accounts for potential errors in the building load prediction model and assesses the planner's ability to handle such variability. As Table~\ref{tab:ablation_study} shows, the total bill is higher than DG-MCTS and LLF. As prediction errors increase, DG-MCTS loses its performance edge over the best heuristic. This effect is stronger for building load predictions—when accuracy drops, performance declines quickly. This may be attributed to the decision tree within MCTS relying heavily on the predicted building load to estimate rewards.
}

\blue{
\label{text:q1b}
\subsection{Effect of Exploration Samples}
We include Table~\ref{table:future_sweep} to illustrate how performance changes with different numbers of exploration samples. We gradually increase the number of exploration samples from 5 to 20 in steps of 5. MCTS-5ES uses 5 samples, MCTS-15ES uses 15, and MCTS-20ES uses 20, where ES stands for exploration steps. The version of DG-MCTS we use for the other experiments has 10 exploration samples and is represented as DG-MCTS (10ES). The table shows that varying the number of exploration samples has minimal impact on overall performance, as the results remain consistent with little deviation. This suggests the chosen exploration samples are sufficient for stable outcomes.
}

%% file: tables/hyper.tex
\begin{table}[h]
\centering
\caption{Hyperparameters for MCTS Configuration}
\label{tab:mcts_hyperparameters}
\begin{tabular}{lcc}
\toprule
\textbf{Hyperparameter} & \textbf{Limits} & \textbf{Value} \\
\midrule
Iterations (Simulation Steps)           &\blue{[$50, 500$]} & 200 \\
Maximum Tree Depth                      &\blue{[$10, 80$]} & 70 \\
Exploration Coefficient ($C$)           &\blue{[$0.5, 2$]} & 1.414 \\
Discount Factor ($\gamma$)              & \blue{[$0.9, 1$]} & 1 \\
Penalty for Missing SOC                 & \blue{[$0.1, 5$]} & 0.5 \\
Penalty for Exceeding Power Gap         & \blue{[$0, 20$]} & 5 \\
Reward for Meeting Required SOC         & \blue{[$0.01, 1$]} & 0.1 \\
Reward for Maximizing SOC               & \blue{[$0.001, 1$]} & 0.01 \\
Estimated Peak Power Adjustment ($\epsilon$) & \blue{[$-10, 10$]} & 0 \\
Exploration Samples (parallel trees)             & \blue{[$5, 20$]} &  10 \\
\bottomrule
\end{tabular}
\label{tab:hyper_conf}
\end{table}

%% file: tables/main_table.tex
\begin{table*}[t]
\centering
\caption{Monthly Total Cost (lower is better).}
\setlength{\tabcolsep}{1.8pt} 
\resizebox{\textwidth}{!}{%
\begin{tabular}{c|c|c|c|c|c|c|c|c}
\toprule
Policy      & MAY          & JUN          & JUL          & AUG          & SEP          & OCT          & NOV          & DEC          \\
\midrule
DG-MCTS        & 5466.96 ± 24.4  & \textbf{6032.16 ± 88.8}  & 6021.98 ± 42.8  & \textbf{8512.24 ± 81.4}  & \textbf{6357.79 ± 36.8}   & \textbf{6744.54 ± 75.7}  & \textbf{5806.73 ± 79.4}  & \textbf{5195.53 ± 127.1} \\
dMCTS       & 5534.54 ± 84.2  & 6050.26 ± 60.0  & 6123.93 ± 73.1  & 8550.6 ± 53.3   & 6467.52 ± 75.9  & 6819.85 ± 62.6  & 5849.69 ± 56.9  & 5317.49 ± 157.0 \\
RL          & \textbf{5416.3 ± 32.8}   & 6067.6 ± 152.1  & \textbf{5913.83 ± 21.4}  & 8571.87 ± 87.9  & 6403.82 ± 105.3 & 6852.93 ± 129.3 & 5898.55 ± 71.9  & 5432.92 ± 135.0 \\
LLF         & 5515.86 ± 30.7  & 6068.35 ± 45.0  & 6045.75 ± 37.1  & 8637.07 ± 44.9  & 6364.37 ± 37.0  & 6802.74 ± 48.8  & 5831.15 ± 33.1  & 5245.1 ± 140.6  \\
EDF         & 5521.67 ± 38.32  & 6076.35 ± 58.8  & 6047.83 ± 37.6  & 8637.67 ± 45.2  & 6369.2 ± 41.1   & 6810.84 ± 51.8  & 5831.53 ± 33.4   & 5292.71 ± 152.6 \\
ReqCharge   & 5577.22 ± 43.8  & 6147.51 ± 47.1  & 6123.04 ± 35.0  & 8743.39 ± 53.7  & 6465.7 ± 46.0   & 6852.65 ± 56.9  & 5939.52 ± 42.2  & 5254.94 ± 65.5  \\
MaxCharge   & 6827.5 ± 188.6  & 7710.79 ± 228.91 & 7577.06 ± 207.8 & 9402.73 ± 144.4 & 8259.19 ± 249.7 & 8348.78 ± 204.7 & 7081.51 ± 184.7  & 7888.98 ± 291.8 \\
\bottomrule
\end{tabular}%
}
\label{table:monthly_policy_costs}
\end{table*}

%% file: tables/missing_soc.tex
\begin{table*}[t]
\centering
\caption{Missing SoC by Policy (lower is better).}
\label{tab:missing_soc}
\begin{tabular}{llllllll}
\toprule
Policy    & DG-MCTS         & dMCTS        & RL           & LLF          & EDF          & ReqCharge    & MaxCharge    \\
\midrule
Mean ± Std & 113.62 ± 83.86 & 164.6 ± 174.24 & 141.92 ± 58.17 & 59.66 ± 84.13 & 58.86 ± 82.29 & 84.87 ± 80.73 & \textbf{27.56 ± 60.55} \\
\bottomrule
\end{tabular}%
\label{table:missing_soc}
\end{table*}

%% file: tables/cars_missing.tex
\begin{table*}[t]
\centering
\caption{Count of Cars with Missing SoC by Policy (lower is better).}
\label{tab:missing_soc_by_policy}
\begin{tabular}{lccccccc}
\toprule
Policy    & DG-MCTS         & dMCTS        & RL           & LLF          & EDF          & ReqCharge    & MaxCharge    \\
\midrule
Mean ± Std & 43.82 ± 30.98 & 73.16 ± 52.88 & 207.34 ± 60.29 & 40.47 ± 30.67 & 41.04 ± 31.36 & 80.57 ± 37.85 & \textbf{4.61 ± 10.22} \\

\bottomrule
\end{tabular}%
\label{table:cars_missing}

\end{table*}

%% file: tables/peak_shaving.tex
\begin{table*}[t]
\centering
\caption{Peak Shaving across all months (lower is better)}
\label{tab:peak_shaving}
\begin{tabular}{lccccccc}
\toprule
Policy       & DG-MCTS          & dMCTS         & RL            & LLF          & EDF          & ReqCharge     & MaxCharge     \\
\midrule
Mean ± Std   & \textbf{-5.17 ± 13.67} & 3.14 ± 11.80  & 18.16 ± 22.81 & 1.77 ± 9.02  & 2.77 ± 10.52 & 11.22 ± 6.15  & 94.02 ± 46.60 \\
\bottomrule
\end{tabular}
\label{table:peak_shaving}
\end{table*}

%% file: tables/ablation.tex
\begin{table*}[t]
\centering
\caption{Ablation study | Sensitivity analysis, on 10 test episodes of August 2023 (lower is better)}
\label{tab:ablation_study}
\begin{tabular}{lcccc|ccc}
\toprule
Month & DG-MCTS             & MCTS/P       & MCTS/H       & MCTS/C  & MCTS(ME) & MCTS(LE) & MCTS(BLF)           \\
\midrule
AUG & \textbf{8512.24 ± 81.42} & 8690.19 ± 86.53 & 8705.76 ± 63.60 & 8521.80 ± 85.13 & 8601.91 ± 52.99 & 8625.59 ± 41.88 & 8670.58 ± 76.77
\\
\bottomrule
\end{tabular}
\label{table:ablation}
\end{table*}

%% file: tables/high_dep_var.tex
\begin{table*}[t]
\centering
\caption{Total cost by policy under a larger departure window (3 hours) (lower is better)}
\label{tab:high_variance}
\begin{tabular}{lcccccc}
\toprule
Month    & DG-MCTS                  & LLF          & EDF          & ReqCharge    & MaxCharge    \\
\midrule
AUG & \textbf{8632.47 ± 73.45} & 8682.06 ± 49.67 & 8682.98 ± 50.74 & 8742.56 ± 55.19 & 9402.73 ± 144.43 \\
\bottomrule
\end{tabular}
\end{table*}

%% file: tables/25cars.tex
\begin{table*}[t]
\centering
\caption{Total Cost of policies under unexpected increases in daily vehicle arrivals (lower is better)}
\begin{tabular}{lcccccc}
\toprule
Month & DG-MCTS & dMCTS & LLF & EDF & ReqCharge & MaxCharge \\
\midrule
AUG & \textbf{8589.0 ± 74.19} & 8650.75 ± 64.98 & 8691.76 ± 53.1 & 8696.32 ± 57.44 & 10345 ± 100 & 11395.53 ± 432.54 \\
\bottomrule
\end{tabular}
\label{tab:25cars}
\end{table*}

%% file: tables/future_sweep.tex
\begin{table*}[t]
\centering
\caption{Total cost and execution time per decision when varying the number of exploration samples (lower is better)}
\begin{tabular}{lccccc}
\toprule
Parameter & DG-MCTS (10ES)             & MCTS-5ES & MCTS-15ES & MCTS-20ES\\
\midrule
Total cost (\$) & 8512.24 ± 79.98 & 8537.95 ± 96.19 & 8511.85 ± 77.3 & 8511.53.13 ± 75.65
\\
Time & 23.75 & 22.37 & 24.39 & 25.01
\\
\bottomrule
\end{tabular}
\label{table:future_sweep}
\end{table*}

%% file: sections/06_conclusion.tex
This paper proposes a Domain-Knowledge Guided Monte Carlo Tree Search (DG-MCTS) approach to address V2B challenges in smart buildings by optimizing charging control. In online decision-making scenarios, we encounter challenges such as the stochastic nature of EV arrivals and departures and limited decision-making time.  To address these challenges, DG-MCTS leverages domain-specific heuristics to guide action selection and prune the action space, enabling it to outperform traditional heuristic algorithms in terms of cost reduction and demand charge optimization within a reasonable computation time. While DG-MCTS shows superior performance, decentralized MCTS (dMCTS) scales more efficiently in scenarios with a larger number of cars and chargers. 
We evaluated both approaches using simulated V2B scenarios with real-world data from Nissan North America, an EV manufacturer and their Silicon Valley lab, a smart building with 15 chargers. The results show that DG-MCTS effectively handles the uncertainty in EV departure times while achieving the lowest monthly total costs and meeting SoC requirements, demonstrating its capability in managing online EV charging under dynamic conditions. Meanwhile, dMCTS provides a scalable alternative that performs well in high-dimensional multi-agent settings.

%% file: sections/09_ack.tex
This material is based upon work sponsored by Nissan Advanced Technology Center-Silicon Valley. Results presented in this paper were obtained using the Chameleon Testbed supported by the National Science Foundation. Any opinions, findings, conclusions, or recommendations expressed in this material are those of the authors and do not necessarily reflect the views of Nissan.

%% file: sections/07_appendix.tex
\input{sections/04a_offline_approach}
 \input{tables/rl_parameters}

\begin{figure}[h]
    \centering
    \includegraphics[width=0.5\linewidth]{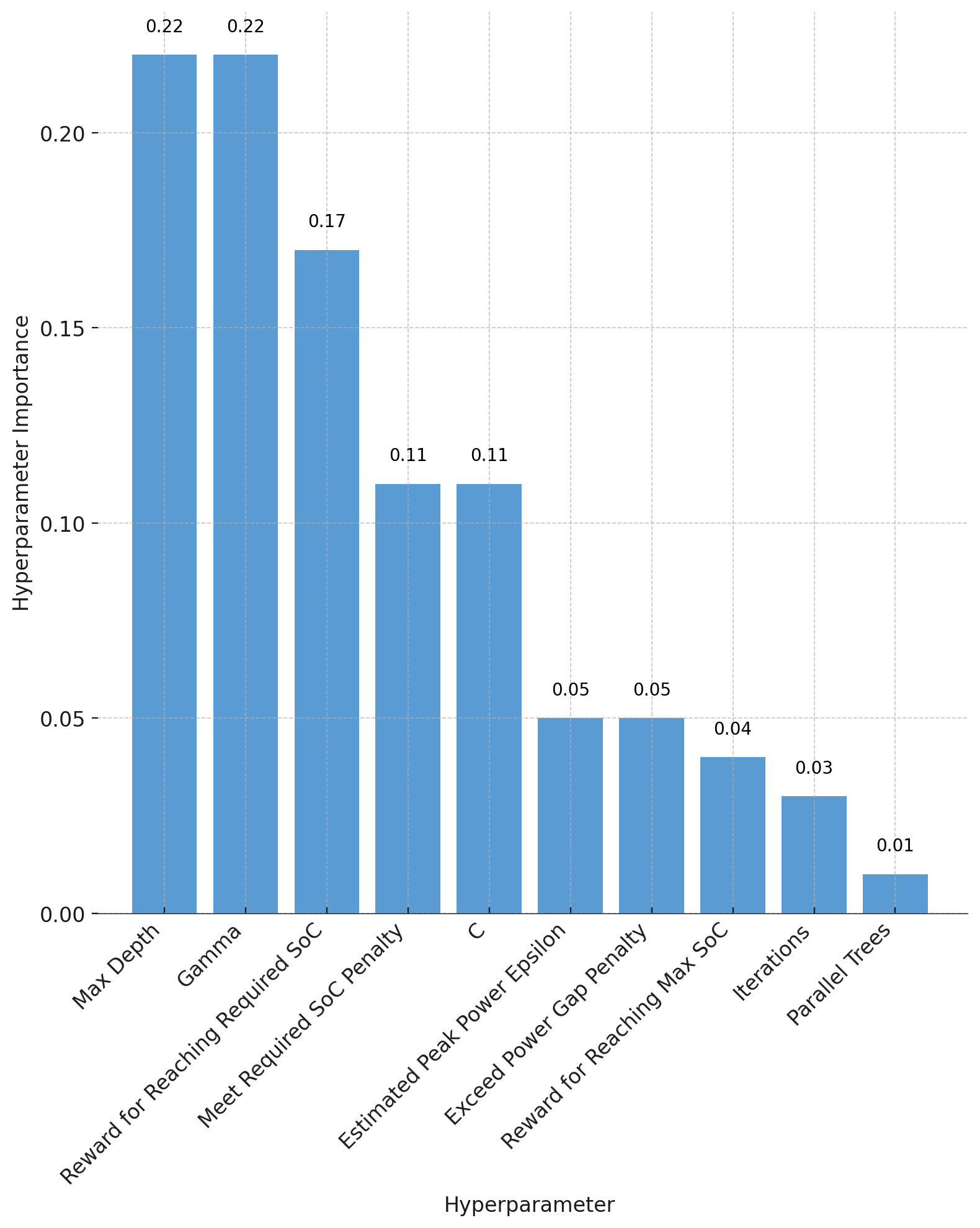}
    \caption{Importance of each factor in hyperparameter exploration}
    \label{fig:optuna}
\end{figure}

%% file: sections/04a_offline_approach.tex
\section{Offline Approach}\label{sec:offline_app}

For any given episode, which is a trajectory of EV arrivals and departures, along with the building load for a single billing period, we can use an offline optimization program to solve the V2B problem and obtain an exact demand cost for that period.
Thus, we formulate a mixed integer linear program (MILP) that can compute the optimal demand cost for any episode.

\subsection{Exact Solution}
The objective of the MILP is \textit{minimizing} the multi-objective weighted sum of the total rewards in Equation~\ref{eq:total_cost}, while meeting the constraints and requirements of the V2B problem. These include ensuring each EV is assigned to a single charger throughout its stay and keeping the action within the charger's limits. 
We use CPLEX~\cite{cplex2009v12} to solve the MILP.

\noindent \textbf{Decision Variables:} 
We use the same set of variables defined in Section~\ref{ssec:problem_spec}. Additionally, $z_v$ is used to represent the gap between the car's required SoC and its SoC at departure.


\noindent \textbf{Constraints:}
To match the car to the charger assignment policy of MCTS (which follows a \textit{first-come first-serve} assignment policy), we pre-compute the car to charger assignments. When $h_{v, k, t} = 1$, it denotes that the assignment of vehicle $v$ to charger $k$ at time $t$, the vehicle stays assigned for the duration of its stay.


We couple the assignment and the charging variables as follows. 
\begin{equation}\label{eq:c3}
   \forall v \in \mathcal{V}, \forall t \in \mathcal{T}: c^v_t \leq \sum_{k} q^{{max}}_{k} \cdot \eta  \cdot h_{v, k, t}
\end{equation}
\begin{equation}\label{eq:c4}
 \forall v \in \mathcal{V}, \forall t \in \mathcal{T}: c^v_t \geq \sum_{k} q^{{min}}_{k} \cdot \eta \cdot h_{v, k, t} 
\end{equation}


We keep track of the EV's SoC before it leaves the charging station, and track if there is any gap to the required SoC.
\begin{equation}\label{eq:soc_req}
    \forall v \in \mathcal{V}: z_v = \left|e^v_{req} - e_{t^v_{end}}^{v} \right|
\end{equation}

Furthermore, for the $i^{th}$ time slot $t_{i}$, for an EV, we can find the amount of energy remaining using $e_{t_{n+1}}^{v}$:
\begin{align}\label{eq:e_s0_new}
\forall v \in \mathcal{V}, t \in \mathcal{T}, 
        \text{ if $i$ = 0 : \quad}
        e_{t_{i}}^{v} = e_{0}^{v} + c_{t_{i}}^{v},
\end{align}
\begin{align}
        \text{\quad otherwise: }
        e_{t_{i}}^{v}= e_{t_{i-1}}^{v} + c_{t_{i}}^{v}
\end{align}
where $e_{0}^{v}$ is the initial charge the EV arrives with.

The demand cost can be computed as (recall that $\Omega$ is the number of aggregate time slots):


\begin{equation}\label{eq:pi_s}
\forall j \in t^{\tau}_j, j \in \{0, 1, \cdots, \Omega \}: \pi_j = \sum_{v \in \mathcal{V}} \sum_{\forall i \in \mathcal{T}^{\tau}} c^v_{t_j} + b^p_{{j}}  
\end{equation}

\begin{equation}\label{eq:c8}
 P_{max} \geq \pi_{j} 
\end{equation}

We need to ensure that a vehicle retains the same charger throughout its stay and do so by maintaining continuous assignment of the car to its charger for its duration of stay: 
\begin{equation}\label{eq:car_charger}
\forall v \in \mathcal{V}, \forall k \in \mathcal{K}, \forall t \in \mathcal{T} \setminus t_{end} : a_{v,k,t} - a_{v,k,t+1} = 0
\end{equation}
\noindent \textbf{Objective} We want to find the minimum energy cost and demand cost, while reducing the gap between the SoC required at departure and the actual departure SoC, given the TOU electricity prices. 
\begin{equation}\label{eq:mod_obj}
    \min \sum_{t\in \mathcal{T}} w_t \cdot \sum_{v \in \mathcal{V}} \left( c^v_t + b^e_t \right) + P_{max} \cdot w^d + w^s \cdot \sum_{v \in \mathcal{V}} z_v
\end{equation}

%% file: tables/rl_parameters.tex
\begin{table}[t]
\centering
\small
\caption{RL (DDPG) Hyperparameters and selected values.}
\begin{tabular}{|>{\raggedright}p{1.7cm}|p{4cm}|p{1.3cm}|}
\hline
Parameter & Description & Range \\ 
\hline
Actor network & Number of units at each layer & [96, 96] \\\hline
Critic network & Number of units at each layer & [96, 96] \\\hline
$\Gamma$ & Discount factor for future reward & 1 \\\hline 
Actor\&Critic learning rate &Learning rate for updating actor and critic networks & $10^{-5}$, $10^{-3}$ \\\hline 
{\it bufferSize} & Batch size for fetching transitions from replay buffer & 64 \\\hline 
{\it batchSize} & Size of the replay buffer & $10^6$ \\\hline 
{\it actionNoise} & Noise added during action exploration & normal(0,0.2)\\\hline 
policyGuideRate & Probability to introduce policy guidance & 0.5 or 0.7 \\\hline  
$\lambda_{S}$, $\lambda_{B}$, $\lambda_{D}$ & Penalty coefficients for SoC requirement, bill cost, and demand charge &1,1, 3\\\hline 
Random seed & Random seed for actor and critic network initialization  &0-5\\\hline  
${\it trainStep}$, ${\it updateStep}$& Training steps and steps per update of target networks& 5, 5\\\hline  

\end{tabular}
\label{tab:hyperparameters}
\end{table}